\documentclass[fleqn,10pt]{wlscirep}
\usepackage[utf8]{inputenc}
\usepackage[T1]{fontenc}

\usepackage{float}     
\usepackage{cleveref}
\usepackage{textcomp} 
\usepackage{enumerate}

\crefname{equation}{equation}{equations}
\crefname{figure}{Fig.}{Figs.}
\crefname{table}{table}{tables}
\Crefname{equation}{Equation}{Equations}
\Crefname{figure}{Figure}{Figures}
\Crefname{table}{Table}{Tables}
\creflabelformat{equation}{#2#1#3}

\newcommand{\um}{\text{\textmu m}}		
\newcommand{\ie}{i.\,e.\ }				
\newcommand{\eg}{e.\,g.\ }				
\newcommand{\ea}{\textit{et al.}}		

\newcommand{\dn}{\Delta n}				
\newcommand{\IT}{\textcolor{black}{T}}			

\newcommand{\IM}{\textcolor{black}{T_{\text{M}}}}			
\newcommand{\dm}{d_{\text{m}}}			
\newcommand{\dM}{d_{\text{M}}}			
\newcommand{\Um}{\mu_{\text{m}}}		
\newcommand{\ic}{\textcolor{black}{T_{{\text{c}}}}}
\newcommand{\dc}{d_{\text{c}}} 			
\newcommand{\Uc}{\mu_{\text{c}}}		

\newcommand{\rmax}{\textcolor{black}{R_{\text{ref}}}}		
\newcommand{\Irmax}{\textcolor{black}{T_{\text{ref}}}}
\newcommand{\rmaxHM}{\textcolor{black}{R_{\text{ref,HM}}}}
\newcommand{\rmaxLM}{\textcolor{black}{R_{\text{ref,LM}}}}
\newcommand{\Prob}{P_{\text{HM}}}
\newcommand{\Iback}{\textcolor{black}{T_{\text{back}}}}
\newcommand{\Ithres}{\textcolor{black}{T_{\text{thres}}}}
\newcommand{\rthres}{\textcolor{black}{R_{\text{thres}}}}

\title{Automated computation of nerve fibre inclinations from 3D polarised light imaging measurements of brain tissue}

\author[1,+,*]{Miriam Menzel}
\author[1,+]{Jan A. Reuter}
\author[1]{David Gräßel}
\author[2,3,4]{Irene Costantini}
\author[1,5]{Katrin Amunts}
\author[1]{Markus Axer} 

\affil[1]{Institute of Neuroscience and Medicine (INM-1), Forschungszentrum Jülich GmbH, Jülich, Germany}
\affil[2]{European Laboratory for Non‑Linear Spectroscopy, University of Florence, Florence, Italy}
\affil[3]{Department of Biology, University of Florence, Florence, Italy}
\affil[4]{National Institute of Optics, National Research Council, Rome, Italy}
\affil[5]{Cécile and Oskar Vogt Institute for Brain Research, University Hospital Düsseldorf, Heinrich Heine University of Düsseldorf, Düsseldorf, Germany}

\affil[*]{Correspondence: m.menzel@fz-juelich.de}
\affil[+]{these authors contributed equally to this work}

\begin{abstract}

The method 3D polarised light imaging (3D-PLI) measures the birefringence of histological brain sections to determine the spatial course of nerve fibres (myelinated axons).
While the in-plane fibre directions can be determined with high accuracy, the
computation of the out-of-plane fibre inclinations is more challenging because they are derived from the \textcolor{black}{amplitude of the birefringence signals}, which depends \eg on the amount of nerve fibres. One possibility to improve the accuracy is to consider the average transmitted light intensity (transmittance weighting). The current procedure requires effortful manual adjustment of parameters and anatomical knowledge. Here, we introduce an automated, optimised computation of the fibre inclinations, allowing for a much faster, reproducible determination of fibre orientations in 3D-PLI. Depending on the degree of myelination, the algorithm uses different models (transmittance-weighted, unweighted, or a linear combination), allowing \textcolor{black}{to} account for regionally specific behaviour. As the algorithm is parallelised and GPU optimised, it can be applied to large data sets. \textcolor{black}{Moreover, it only} uses images from standard 3D-PLI measurements \textcolor{black}{without tilting, and can therefore be applied to existing data sets} from previous measurements. The functionality is demonstrated on unstained coronal and sagittal histological sections of vervet monkey and rat brains. 

\end{abstract}
\begin{document}

\flushbottom
\maketitle

\thispagestyle{empty}


\section*{Introduction}

To better understand the function of the brain and to treat neurological diseases, a detailed reconstruction of the intricate and densely grown nerve fibre network is needed.
\textcolor{black}{Diffusion magnetic resonance imaging allows to study the course of nerve fibre pathways in vivo and for whole brain volumes\cite{mori2006,tuch2003}, but the resolution is not sufficient to reconstruct individual nerve fibres\cite{hagman2003,chang2015}. Microscopy techniques like optical coherence tomography\cite{men2016,magnain2015}, light-sheet microscopy\cite{mertz2010,stefaniuk2016}, or two-photon fluorescence microscopy\cite{silvestri2014,amato2016} allow an in-depth scan of the sample and provide microscopic resolution in all three dimensions. However, they can only be applied to smaller brain volumes and need advanced image processing to follow the course of nerve fibres, which is especially challenging in regions with densely packed fibres. 
The microscopy technique \textit{3D polarised light imaging} (3D-PLI)\cite{MAxer2011_1,MAxer2011_2} analyses the spatial course of nerve fibre pathways by transmitting polarised light through unstained histological brain sections and measuring their \textit{birefringence} (optical anisotropy).
In contrast to the other techniques, 3D-PLI requires sectioning of the brain and advanced registration algorithms to recover the original brain volume. But it allows to retrieve the three-dimensional orientations of the nerve fibres also in densely grown regions and for large brain sections, making 3D-PLI a powerful approach for analysing the brain's nerve fibre network in microscopic detail.}

\textcolor{black}{The three-dimensional nerve fibre orientations are derived from the measured birefringence signals, which are mainly caused by \textit{myelin} -- a stack of cell membranes with highly ordered molecular structure that surrounds} most axons in the white matter\cite{zilles2010}. (In the following, the term \textit{nerve fibre} will solely be used to refer to myelinated axons.)
The \textcolor{black}{orientation} of the nerve fibres \textcolor{black}{within the section plane} can be determined with high degree of accuracy, provided the fibre bundles have a well-defined orientation\cite{menzel2021}. The computation of the \textcolor{black}{out-of-plane fibre orientation (\textit{inclination})} requires a more sophisticated analysis:
\textcolor{black}{The} fibre inclination is \textcolor{black}{directly} related to the \textcolor{black}{amplitude of the birefringence signal}\textcolor{black}{, which} also depends on other factors such as the amount of nerve fibres\cite{dohmen2015}.
In order to determine the fibre inclinations independent of the tissue composition, a tiltable specimen stage can be used: By repeating the 3D-PLI measurement for different tilting angles, it is possible to change the perspective onto the nerve fibres in a predictable manner and reconstruct the fibre inclinations. \textcolor{black}{Tilting has successfully been employed to improve the determined inclination angles for 3D-PLI measurements with in-plane resolutions comparable to the brain section thickness\cite{schmitz2018}. However, analysing images obtained from tilting microscopes with higher in-plane resolution remains a challenge. Moreover, many} high-resolution polarising microscopes are not equipped with a tilting stage. 

An alternative way to adjust the determined inclination angles is to use the transmittance (polarisation-independent transmitted light intensity) as a measure of the amount of nerve fibres in the brain section (\textit{transmittance weighting})\cite{reckfort}: Due to its highly layered structure, myelin has a large scattering coefficient\cite{schwarzmaier1997} so that regions with a high density of nerve fibres (myelinated axons) appear dark in the transmittance image. 

\textcolor{black}{Transmittance weighting has the advantage that it can be applied to (existing) data sets from standard 3D-PLI measurements without tilting. However, using the transmittance to adjust the computed fibre inclinations poses several challenges:}
Previous work of our group\cite{menzel2020} has shown that the transmittance is not solely a measure of the myelin density, but also depends on the inclination of the nerve fibres: regions with out-of-plane nerve fibres show a lower transmittance than in-plane nerve fibres, even if they contain the same amount of nerve fibres. 
In addition, the transmittance can only serve as a measure of the myelin density (amount of nerve fibres) if the attenuation of myelin is dominant compared to the attenuation of other tissue components.
Apart from myelinated axons, brain tissue contains unmyelinated axons, neuronal cell bodies, dendrites, synapses, glial cells, and blood capillaries, which are all differently distributed and have locally variable attenuation properties\cite{schuenke2007,zilles2010}.
Therefore, the transmittance cannot serve as a measure of myelination in regions with a low amount of myelinated axons\textcolor{black}{.}

Up to now, the nerve fibre inclinations were computed using a transmittance-weighted model for the whole brain section, requiring manual adjustment of the resulting parameters and advanced anatomical knowledge to account for different tissue compositions.
Based on transmittance and retardation histograms, an expert needed to iteratively adjust various threshold parameters to identify the optimum set of parameters that yields particular fibre inclinations in anatomically known brain regions while reducing the number of regions with under/over-estimated fibre inclinations\textcolor{black}{.}

Here, we introduce a fully automated computation of the nerve fibre inclinations, taking the inclination-dependence of the transmittance and different degrees of myelination into account. 
Instead of applying the transmittance-weighted model to the whole brain section, we use a regionally specific computation of the fibre inclinations: The transmittance-weighted model is only used in regions with a high degree of myelination. In regions with a low degree of myelination, an unweighted model is used. The separation of regions with low and high degrees of myelination is based on an algorithm proposed by Benning \ea\cite{benning2021}. However, instead of using a binary classification, we introduce the important concept of transition zones, allowing to consider subtle changes in the tissue composition and to compute more accurate fibre inclinations at boundaries.
As the automated computation does not require any manual adjustment of the parameters, it allows for a much faster and reproducible computation of the three-dimensional nerve fibre orientations for 3D-PLI measurements of brain tissue.


\section*{Methods}
\label{methods}

\subsection*{Experimental methods}

\textcolor{black}{We demonstrate the functionality of our algorithm on coronal and sagittal sections of primate and rodent brains. To validate the automatically computed nerve fibre inclinations, we compared our results from 3D-PLI to in-depth-tissue scans from two-photon fluorescence microscopy.}


\paragraph{Preparation of brain sections.} 
The brain sections were obtained from three healthy rats (Wistar, male, three months old) and two healthy vervet monkeys (male, between one and two years old).
All animal procedures were approved by the institutional animal welfare committee at Forschungszentrum Jülich GmbH, Germany, and were in accordance with the European Union and National Institutes of Health guidelines for the use and care of laboratory animals and in compliance with the ARRIVE guidelines. Euthanasia of rats was carried out under controlled isoflurane inhalation followed by decapitation.
The monkey brains were obtained in accordance with the Wake Forest Institutional Animal Care and Use Committee (IACUC \#A11-219). Euthanasia procedures conformed to the AVMA Guidelines for the Euthanasia of Animals, using ketamine/pentobarbital anesthesia followed by perfusion with phosphate buffered saline and fixation with 4\,\% paraformaldehyde.

The brains were removed from the skull within 24 hours after death, fixed with 4\,\% buffered formaldehyde for several weeks, cryo-protected with 20\,\% glycerine and 2\,\% dimethyl sulfoxide, deeply frozen, and coronally or sagittally cut with a cryostat microtome (Polycut CM 3500, Leica, Microsystems, Germany) into sections of 60\,\um. A coronal vervet monkey brain section (no.\ 512), a sagittal vervet monkey brain section (no.\ 374), a coronal rat brain section (no.\ 185), and a sagittal rat brain section (no.\ 176) were selected for evaluation. The sections were mounted on glass slides, embedded in 20\,\% glycerine solution, cover-slipped, sealed with lacquer, and measured with 3D-PLI afterwards. 

\begin{figure}[h]
	\centering
	\includegraphics[width=0.8\textwidth]{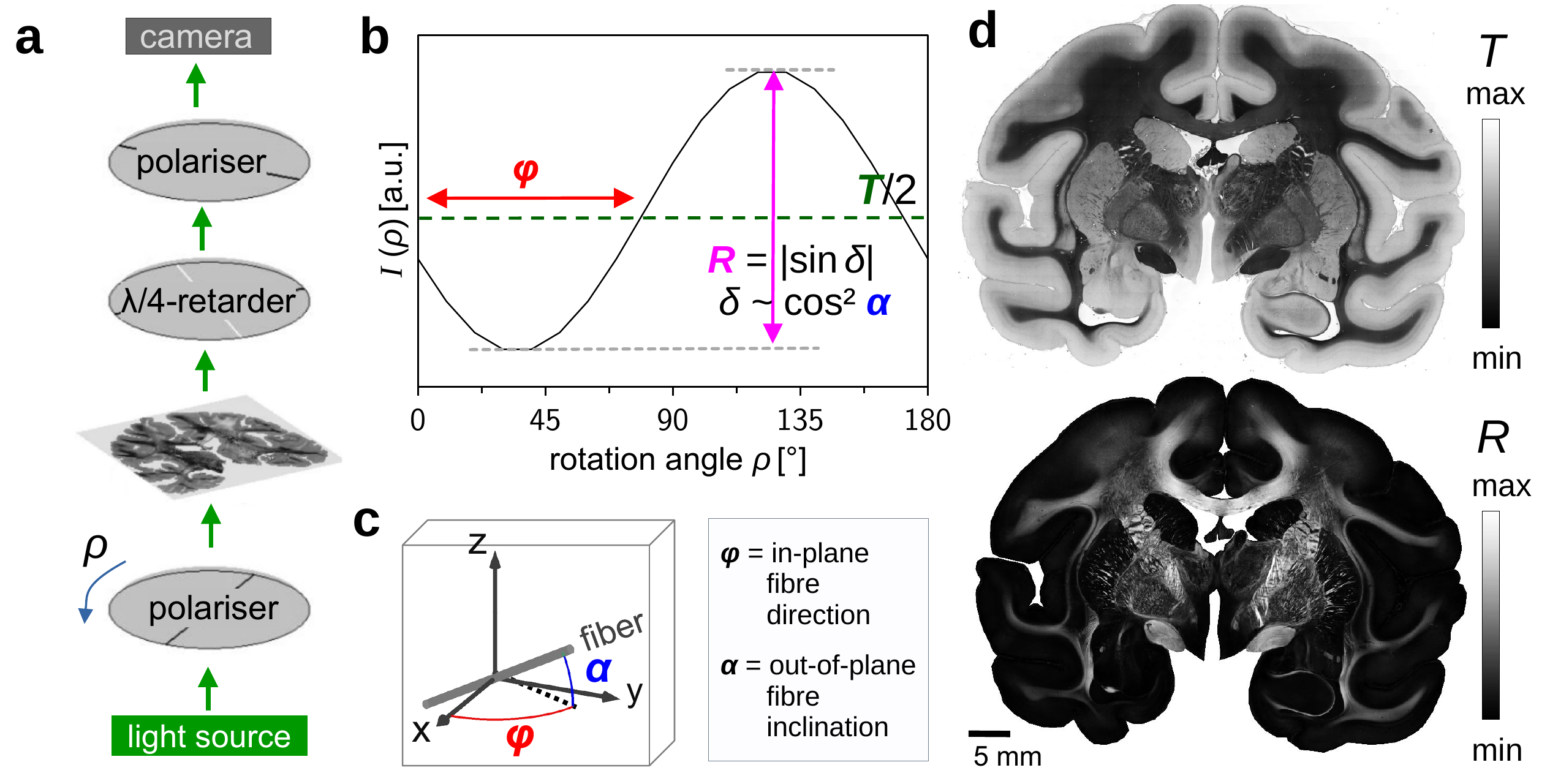}
	\caption{\textcolor{black}{3D-PLI measurement and analysis. (\textbf{a}) Schematic drawing of the set-up (not to scale). (\textbf{b}) Transmitted light intensity of one image pixel for different rotation angles $\rho$ of the polariser. The transmittance ($\textcolor{black}{T}$) and retardation ($\textcolor{black}{R}$) are obtained from the average and normalized amplitude of the signal, respectively. (\textbf{c}) Definition of the in-plane fibre direction $\varphi$ and the out-of-plane fibre inclination $\alpha$, which are derived from the phase and amplitude of the signal, respectively, as indicated in (b). (\textbf{d}) Transmittance image (top) and retardation image (bottom) for the coronal vervet monkey brain section.}}
	\label{fig:1}
\end{figure}

\paragraph{\textcolor{black}{3D polarized light imaging (3D-PLI).}}
The 3D-PLI measurements were performed with the polarising microscope (LMP-1, Taorad GmbH, Germany), as described in Menzel \ea\ \cite{menzel2021,menzel2020}: Incoherent light with a wavelength of about 550\,nm was transmitted through a linear polariser, the brain section, and a circular analyser \textcolor{black}{(quarter-wave retarder and linear polariser), see \cref{fig:1}a}. The \textcolor{black}{first} polariser was rotated in steps of $10^{\circ}$ and for each rotation angle ($\rho = \{0^{\circ}, 10^{\circ}, \dots, 170^{\circ}\}$) an image was recorded with a CCD camera (Qimaging Retiga 4000R), yielding a pixel size in object space of about 1.33\,\um. For each image pixel, the intensity values of the resulting image series form a sinusoidal signal with respect to the rotation angle \textcolor{black}{(see \cref{fig:1}b)}, which can be described as\cite{MAxer2011_1,MAxer2011_2}:
\begin{align}
I(\rho) = \frac{\IT}{2} \Big(1 + \sin\delta \, \sin\big(2(\rho-\varphi)\big) \Big).
\end{align}
The average of the signal (\textit{transmittance} $\IT$) is a measure of the tissue attenuation, caused by absorption and scattering. 
The phase of the signal (\textit{direction} $\varphi$) indicates the in-plane direction angle of the nerve fibres \textcolor{black}{within the brain section}.
The amplitude of the normalised signal (\textit{retardation} $\textcolor{black}{R} = \vert\sin\delta\vert$) indicates the strength of birefringence and is related to the out-of-plane angle of the nerve fibres (\textit{inclination} $\alpha$) \textcolor{black}{via $\delta \sim \cos\alpha^2$ (see below).} \textcolor{black}{\Cref{fig:1}c illustrates the definition of $\varphi$ and $\alpha$, \cref{fig:1}d shows an example of the resulting transmittance and retardation images.} 

\paragraph{\textcolor{black}{Two-photon fluorescence microscopy (TPFM).}}

\textcolor{black}{To validate the automatically computed nerve fibre inclinations, we compared the results from 3D-PLI to two-photon fluorescence microscopy (TPFM) measurements of the same brain section.}
The TPFM measurements were performed on a coronal rat brain section ($6 \times 8$ tiles in the caudate putamen, cf.\ \cref{fig:6}b) \textcolor{black}{as} described in Menzel \ea\cite{menzel2020}, with a custom-made two-photon fluorescence microscope \cite{silvestri2014,costantini2017,costantini2021} consisting of a model-locked titanium-sapphire laser with a wavelength of 800\,nm and a water-immersion $25\times$ objective lens (LD LCI Plan-Apochromat 25x/0.8 Imm Corr DIC M27), achieving a resolution of $0.244 \times 0.244 \times 1$\,\um$^3$. The fluorescence signals were collected by two photomultiplier tubes, detecting red and green fluorescence, respectively.
TPFM allows a full in-depth scan of the 60\,\um\ thin brain section, yielding one image every $z = 1$\,\um. The axial displacement was realised with a closed-loop piezoelectric stage. A motorised xy-stage enabled tile-wise scanning of the sample. The sample was measured in tiles of $250 \times 250$\,\um$^2$, with an overlap of 10\,\%, and the resulting images were stitched together using the software \textit{TeraStitcher}\cite{bria2012}. 
Due to the slightly different autofluorescence signal of different tissue components, the nerve fibre bundles can be manually traced across different slices of a TPFM image stack. To determine the inclination angle of a fibre bundle, the cross-sections of the fibre bundle were determined in the first and last slices of the image stack that still show the fibre bundle. The inclination angle $\alpha$ was computed geometrically from the midpoint positions of the cross-sections, taking the number of images (z-depth) between the first and last slice into account\textcolor{black}{.}
230 fibre bundles in the caudate putamen were selected for evaluation. Only well-separated fibre bundles with well-defined fibre inclination were chosen. 
To enable a comparison between the geometrically computed TPFM inclinations and the automatically computed 3D-PLI inclinations, the transmittance image of the 3D-PLI measurement was registered onto the maximum intensity projection of the TPFM image stack using affine transformations, and the individual fibre bundles were masked in the registered transmittance image. To avoid registration artefacts, the inverse transformation was applied to the resulting fibre mask, allowing to locate the fibre bundles in the original (non-registered) inclination image \textcolor{black}{and compare the 3D-PLI inclination to the corresponding TPFM inclination.}
To ensure that the tissue had not deformed between measurements, the maximum intensity projection of the TPFM image stack was compared to the transmittance image, ensuring that visible borders of nerve fibre bundles overlap.


\subsection*{\textcolor{black}{Models for computing the fibre inclination angle in 3D-PLI}}


\textcolor{black}{The out-of-plane inclination angle $\alpha$ of the nerve fibres is related to the \textcolor{black}{amplitude of the 3D-PLI signal (retardation $\textcolor{black}{R}$)}} via \cite{menzel2015}:
\begin{align}
\textcolor{black}{R} = |\sin\delta| \approx \left| \sin \left( \frac{2 \pi}{\lambda} \, \dn \, \dm \, \cos^2 \alpha \right) \right|, \label{eq:ret}
\end{align}
where $\lambda$ is the wavelength of the incident light, $\dm$ the thickness of the birefringent tissue (myelin), and $\dn$ the birefringence of the material\textcolor{black}{. As these parameters} are usually not precisely known \cite{MAxer2011_1}\textcolor{black}{, the fibre inclinations can only be computed based on certain assumptions.}

\paragraph{Unweighted model.}
	Assuming that $\dn\,\dm$ does not change much within the measured brain section, the fibre inclination angle can be computed for each image pixel via:
	\begin{align}
	\alpha = \arccos \left( \sqrt{\frac{\arcsin(\textcolor{black}{R})}{\arcsin(\rmax)}}\right), \label{eq:alpha}
	\end{align}
where $\rmax$ is the \textcolor{black}{retardation} expected to occur in regions with densely packed, parallel in-plane nerve fibres ($\rmax \equiv \textcolor{black}{R}_{\alpha = 0^{\circ}}$)\textcolor{black}{.}
	
\paragraph{Transmittance-weighted model.} Usually, the amount of nerve fibres (and thus the amount of birefringent material $\dm$) changes notably within a brain section. \textcolor{black}{When a region with a large amount of nerve fibres is used as reference, the unweighted model leads to} an over-estimation of the fibre inclination angles in \textcolor{black}{other} regions with a \textcolor{black}{smaller} amount of nerve fibres. To take the different amounts of nerve fibres and hence the different retardation values into account, the transmittance \textcolor{black}{can be} used as a reference. Due to the high scattering coefficient of myelin, regions with many nerve fibres appear darker than regions with less nerve fibres. When light with intensity $I_0$ is transmitted through a material with thickness $d$ and attenuation coefficient $\mu$, the resulting light intensity $T$ is attenuated according to the Lambert-Beer law:
	\begin{align}
	\IT = I_0 \, \exp\left(-\mu d\right) \quad \Leftrightarrow \quad \mu d = \ln\left(\frac{I_0}{\IT}\right). \label{eq:LB}
	\end{align}
	
In brain tissue, light passes through birefringent tissue (myelin) with attenuation coefficient $\Um$, and non-birefringent tissue (\eg cell bodies) with attenuation coefficient $\Uc$. 
For now, we assume that both attenuation coefficients do not depend on the exact tissue composition or the fibre inclination. 
The maximum thickness of birefringent tissue in the brain section is defined as $\dM$, and the transmittance in this region as: $\IM = \tilde{I}_0 \, \exp(-\Um \dM)$, where $\tilde{I}_0$ is the transmitted light intensity taking the non-birefringent layer with thickness $(d - \dM)$ into account: $\tilde{I}_0 = I_0 \, \exp(-\Uc (d-\dM))$. Equivalently, we define the transmittance in a region that is completely filled with non-birefringent tissue as: $\ic = \tilde{I}_0 \, \exp(-\Uc \dM)$. The transmittance in any other brain region is then defined as: $\IT = \tilde{I}_0 \, \exp(-\mu \dM) = \tilde{I}_0 \, \exp(-\Um \dm) \exp(-\Uc \dc)$, with $\dM = \dm + \dc$ and $\dm$ being the local thickness of birefringent tissue. With these definitions, we can write:
	\begin{align}
	\mu \dM &= \Um\dm + \Uc\dc = \Um\dm + \Uc (\dM - \dm) \label{eq:ut} \\
	\Leftrightarrow \dm &\overset{(\ref{eq:ut})}{=} \dM \frac{\mu - \Uc}{\Um - \Uc} \overset{(\ref{eq:LB})}{=} \dM \frac{\ln \big( \ic / \IT \big)}{\ln \big( \ic / \IM \big)}.
	\end{align}
	Inserting this into \cref{eq:ret}, yields a modified formula for the fibre inclination angle \cite{reckfort}:
	\begin{align}
	\alpha = \arccos \left( \sqrt{\frac{\arcsin(\textcolor{black}{R})}{\arcsin(\rmax)} \cdot \frac{\ln \big( \ic / \IM \big)}{\ln \big( \ic / \IT \big)}}\,\right),\label{eq:alpha2}
	\end{align}
	where $\rmax = |\sin(2\pi \, \dn\,\dM/\lambda)|$ is the retardation of a region with parallel in-plane nerve fibres with $\dm = \dM$ and $\alpha=0^{\circ}$. 
	
\paragraph{\textcolor{black}{Combined model.}}

\textcolor{black}{As mentioned in the Introduction, the transmittance is only a good measure of the amount of nerve fibres (myelinated axons) if the attenuation of myelin is dominant compared to the attenuation of other tissue components.
In regions with a high degree of myelination (\textit{HM-regions}), the attenuation can be considered to be mostly caused by the scattering of myelin so that the fibre inclinations can be computed with the transmittance-weighted model. However, in regions with a low degree of myelination (\textit{LM-regions}), the transmittance is dominated by other tissue components such as cell bodies so that the unweighted model should be used instead to compute the fibre inclinations.} 

\textcolor{black}{To account for different tissue compositions, we defined $\Prob \in [0,1]$ as} the probability that a region is highly myelinated \textcolor{black}{and computed the nerve fibre inclinations using a linear combination of both models,} the transmittance-weighted model for HM-regions (\cref{eq:alpha2}) and the unweighted model for LM-regions (\cref{eq:alpha}):

\begin{align}
\alpha &= \Prob \cdot \alpha_{\text{HM}} + (1 - \Prob) \cdot \alpha_{\text{LM}} \notag \\
&= \Prob \cdot \arccos\left(\sqrt{\frac{\arcsin(\textcolor{black}{R})}{\arcsin(\rmaxHM)} \cdot \frac{\ln(\ic / \IM)}{\ln(\ic / \IT)}}\right)
+
(1-\Prob)\cdot\arccos \left( \sqrt{\frac{\arcsin(\textcolor{black}{R})}{\arcsin(\rmax^*)}}\right),
\label{eq:final}
\end{align}
where
\begin{align}
\rmax^* &\equiv \Prob \cdot \rmaxHM + (1-\Prob) \cdot \rmaxLM.
\end{align}
\textcolor{black}{In this combined model, the $\rmax$ value was defined separately for HM- and LM-regions ($\rmaxHM$, $\rmaxLM$), and  a linearly interpolated value $\rmax^*$ was used in the transition zone between HM- and LM-regions, preventing an over-estimation of fibre inclinations in regions with small amounts of nerve fibres.}


\subsection*{Determination of the degree of myelination}

\textcolor{black}{To determine the degree of myelination ($\Prob$) for all image pixels of a measured brain section, we first used a binary classification to separate LM-regions ($\Prob=0$) from HM-regions ($\Prob=1$). Based on this initial separation, we then defined transition zones with $0 < \Prob < 1$.}


\paragraph{Binary classification.}

\textcolor{black}{We followed} the classification \textcolor{black}{introduced in} Benning \ea\cite{benning2021} to separate regions with low and high myelination (LM- and HM-regions). Note that \textcolor{black}{the authors used the terms} ``grey matter'' and ``white matter''\textcolor{black}{, while we use the terms LM- and HM-regions} to avoid confusion with the anatomical definitions.
The separation of regions was performed \textcolor{black}{by analysing the transmittance ($\textcolor{black}{T}$) and retardation ($\textcolor{black}{R}$) images from 3D-PLI.}

\begin{figure}[h]
	\centering
	\includegraphics[width=0.9\textwidth]{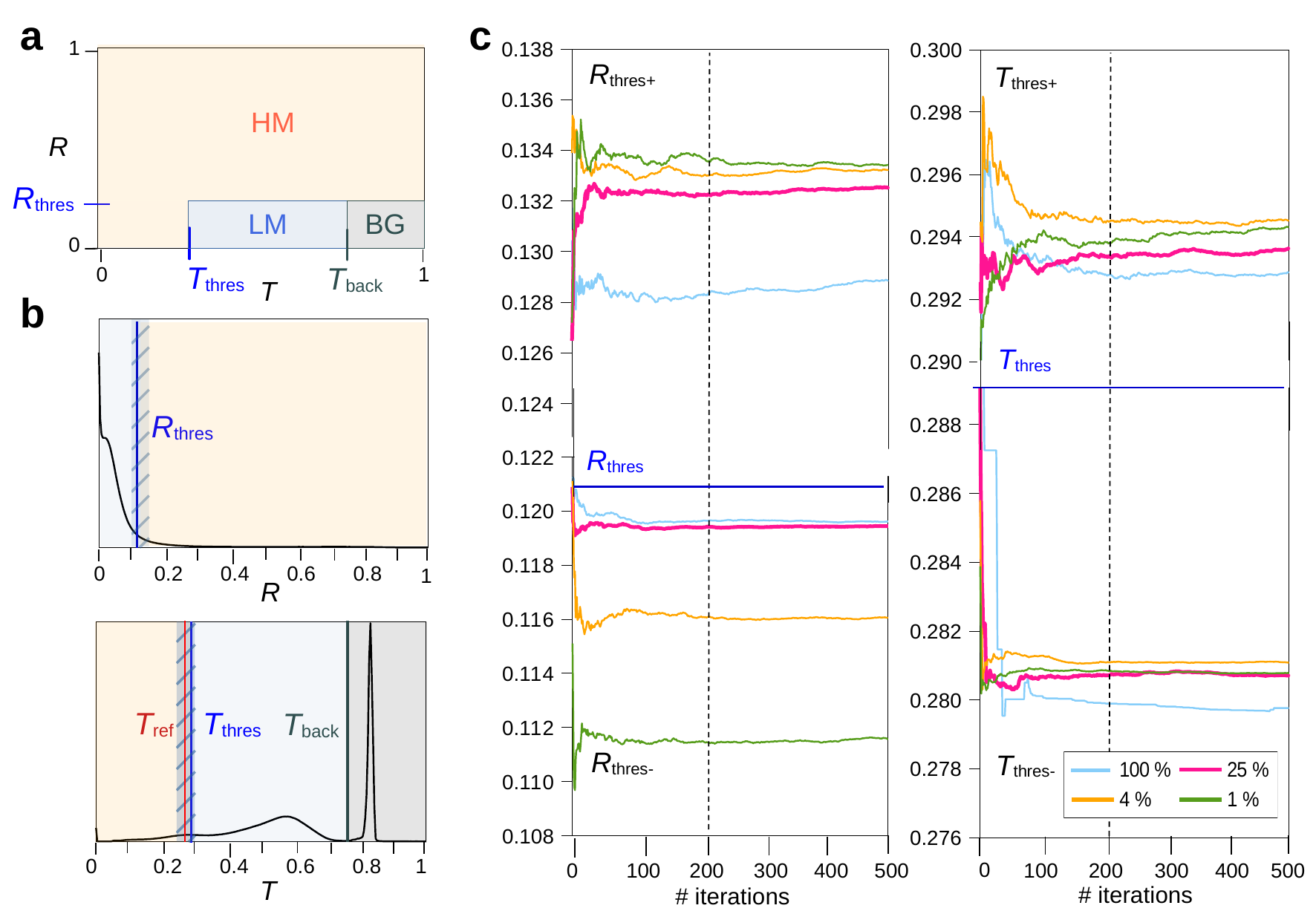}
	\caption{Optimum choice of iteration number and sample size when computing the HM-probability map, shown exemplary for a coronal rat brain section (cf.\ \cref{fig:6}b). \textcolor{black}{(\textbf{a}) Separation of HM-regions (orange), LM-regions (blue), and BG-regions (grey) based on the retardation $\textcolor{black}{R}$ and transmittance $T$ values of the respective image pixel. Depending on whether the values lie above or below $\rthres$, $\Ithres$, or $\Iback$, the pixel is assigned to one or the other region.} (\textbf{b}) Retardation and transmittance histograms (256 bins) \textcolor{black}{of the coronal rat brain section}. The parameters $\rthres$, $\Irmax$, $\Ithres$, and $\Iback$ are indicated by vertical solid lines. When computing the histograms from different sets of image pixels, the threshold parameters $\rthres$ and $\Ithres$ may vary, which is indicated by striped areas \textcolor{black}{(the striped areas are only shown for illustration purposes and depend on the respective set of image pixels).} (\textbf{c}) \textcolor{black}{Corresponding positive/negative means ($R_{\text{thres}+/-}$ and $T_{\text{thres}+/-}$)}, computed for up to 500 iterations and different sample sizes (100\,\%, 25\,\%, 4\,\%, 1\,\% of all image pixels).}
	\label{fig:2}
\end{figure}

While all regions with high birefringence ($\textcolor{black}{R} > \rthres$) can be considered as HM-regions, not all regions with low birefringence ($\textcolor{black}{R} \leq \rthres$) can be considered as LM-regions because regions with crossing fibres, out-of-plane fibres, or a smaller number of nerve fibres also result in smaller birefringence values.
As in-plane crossing and out-of-plane fibres have similar or even lower transmittance values than regions with \textcolor{black}{parallel} in-plane fibres\cite{menzel2020}, the transmittance in a region with many \textcolor{black}{parallel} in-plane fibres can be used as a reference ($\Irmax$): Hence, HM-regions are defined by regions with similar or lower transmittance values ($\IT \lesssim \Irmax$), while LM-regions are restricted to regions with higher transmittance values ($\IT > \Irmax$), and \textcolor{black}{BG-regions (background }not containing any tissue) to regions with much higher transmittance values ($\IT \gg \Irmax$). 
For better separation, \textcolor{black}{the threshold parameters $\Ithres (\sim \Irmax)$ and $\Iback (\gg \Irmax)$} were used \textcolor{black}{instead. This yields the following classification for regions with high/low myelination and background (see \cref{fig:2}a):} 

\begin{itemize}[noitemsep]
    \item HM-regions (high myelination): $(\textcolor{black}{R} > \rthres) \vee (\IT < \Ithres)$,
    \item LM-regions (low myelination):\,\, $(\textcolor{black}{R} \leq \rthres) \wedge (\Ithres \leq \IT \leq \Iback)$,
    \item \textcolor{black}{BG-regions} (no tissue): \quad\quad\quad\, $(\textcolor{black}{R} \leq \rthres) \wedge (\IT > \Iback)$.
\end{itemize}

\textcolor{black}{All three threshold parameters ($\rthres$, $\Ithres$, $\Iback$) were computed from} points of maximum curvature in the \textcolor{black}{retardation and transmittance histograms (see blue/black vertical lines in \cref{fig:2}b)}.
To prevent the background from having any influence on the computed fibre inclinations, the retardation and transmittance of all background pixels were set to their minimum and maximum value, respectively, before analysing the corresponding histograms.
Before computing the transmittance histogram, the values were normalised to [0,1] and a circular median filter with 5\,px radius was applied to the transmittance image. We selected a radius of 5\,px (and not 10\,px as in Benning \ea) because a larger radius leads to more clouding artefacts in the resulting inclination map, while a smaller radius leads to increased noise and makes non-relevant details like cells more visible.

As in Benning \ea\cite{benning2021}, the threshold parameters $\rthres$ and $\Iback$ were computed as the points of maximum curvature behind and before the biggest peak in the retardation and transmittance histogram, respectively. In contrast to Benning \ea\cite{benning2021} who used 128 bins, we decided to use histograms with 256 bins in order to increase accuracy.
However, histograms with 256 bins might show several small peaks or a plateau (\textcolor{black}{cf.\ \cref{fig:2}b on the very left}), which makes the determination of the threshold parameters difficult. To ensure that no outliers were chosen, we first considered a histogram with 64 bins and computed the threshold parameters (points of maximum curvature) in a range corresponding to the full-width at half-maximum (FWHM) of the peaks, using larger ranges as Benning \ea\cite{benning2021} to account for thinner peaks: 20 $\times$ FWHM behind the retardation peak and 10 $\times$ FWHM before the transmittance peak. Then, we considered histograms with 128 bins and recomputed the threshold parameters in a range of bins close to the previous results [$2 \cdot (\text{result} - 1), 2 \cdot (\text{result} + 1)$]. The last step was repeated for histograms with 256 bins.
In this way, the computed threshold parameters stay close to the previous results while being more accurate. 

The threshold parameter $\Ithres$ was computed as the point of maximum curvature between $\Irmax$ and $\Iback$\textcolor{black}{. The value for $\Irmax$ was obtained by computing} the average transmittance of the region \textcolor{black}{with the \textcolor{black}{highest retardation values}, which is expected to contain the highest amount of parallel in-plane fibres (cf.\ \cref{eq:ret}). To avoid outliers, the union-find connected components algorithm by Oliveira \ea \cite{Oliveira2010ASO} was used to identify a region of connected pixels (0.009--0.011\,\% of the image size) that has the highest retardation values.}

\paragraph{\textcolor{black}{Transition zones.}}

When the inclination angles \textcolor{black}{are computed separately} for HM- and LM-regions, artificially sharp borders \textcolor{black}{might occur}, \eg in regions where nerve fibres from the white matter spread out into the cortex.
To account for regions that do not clearly belong to one or the other tissue type, \textcolor{black}{we defined} \textit{transition zones} between HM- and LM-regions.

To determine the width of these transition zones, the threshold parameters $\rthres$ and $\Ithres$ were considered, which have a major impact on the separation of HM- and LM-regions (cf.\ striped \textcolor{black}{blue} areas in \cref{fig:2}b): 
In a transition zone, slight changes of the threshold parameters cause the region to be identified as a different tissue type.
To find out whether a separation is susceptible to small changes, the threshold parameters were recomputed several times from slightly different retardation and transmittance histograms that were generated by \textit{bootstrapping} (random sampling with replacement): Pixels were randomly drawn from the retardation and transmittance images; the same pixel could be selected multiple times. 
To identify the best compromise between computing time and accuracy, the algorithm was run for different sample sizes (100\,\%, 25\,\%, 4\,\%, 1\,\% of all image pixels) for 500 iterations. 
From the resulting values for $\rthres$ and $\Ithres$, the \textit{positive/negative means} ($\textcolor{black}{R_{\text{thres}+/-}}$ and $\textcolor{black}{T_{\text{thres}+/-}}$) were computed, respectively\textcolor{black}{. The positive (negative) mean is the mean of all recomputed threshold values that are larger (smaller) than the original threshold value.} The whole process was repeated 10 times and the average values for each iteration were forwarded to further analysis. 

\Cref{fig:2}c shows the positive and negative means for the different sample sizes and up to 500 iterations for a coronal rat brain section. The different sample sizes yield similar curves. The smaller the sample size, the more the values differ from the ones obtained with 100\,\% sample size.  
After 200 iterations (vertical dashed line), the values for 100\,\% and 25\,\% sample size (cyan and magenta curves) do not change much and the maximum difference between 25\,\% and 100\,\% sample size is still within one histogram bin ($1/256=0.0039$). Therefore, a sample size of 25\,\% and 200 iterations were selected for computing $\textcolor{black}{R_{\text{thres}+/-}}$ and $\textcolor{black}{T_{\text{thres}+/-}}$\textcolor{black}{, which define the width of the transition zone between LM- and HM-regions}.

\paragraph{HM-probability map.}

\begin{figure}[h]
	\centering
	\includegraphics[width=0.9\textwidth]{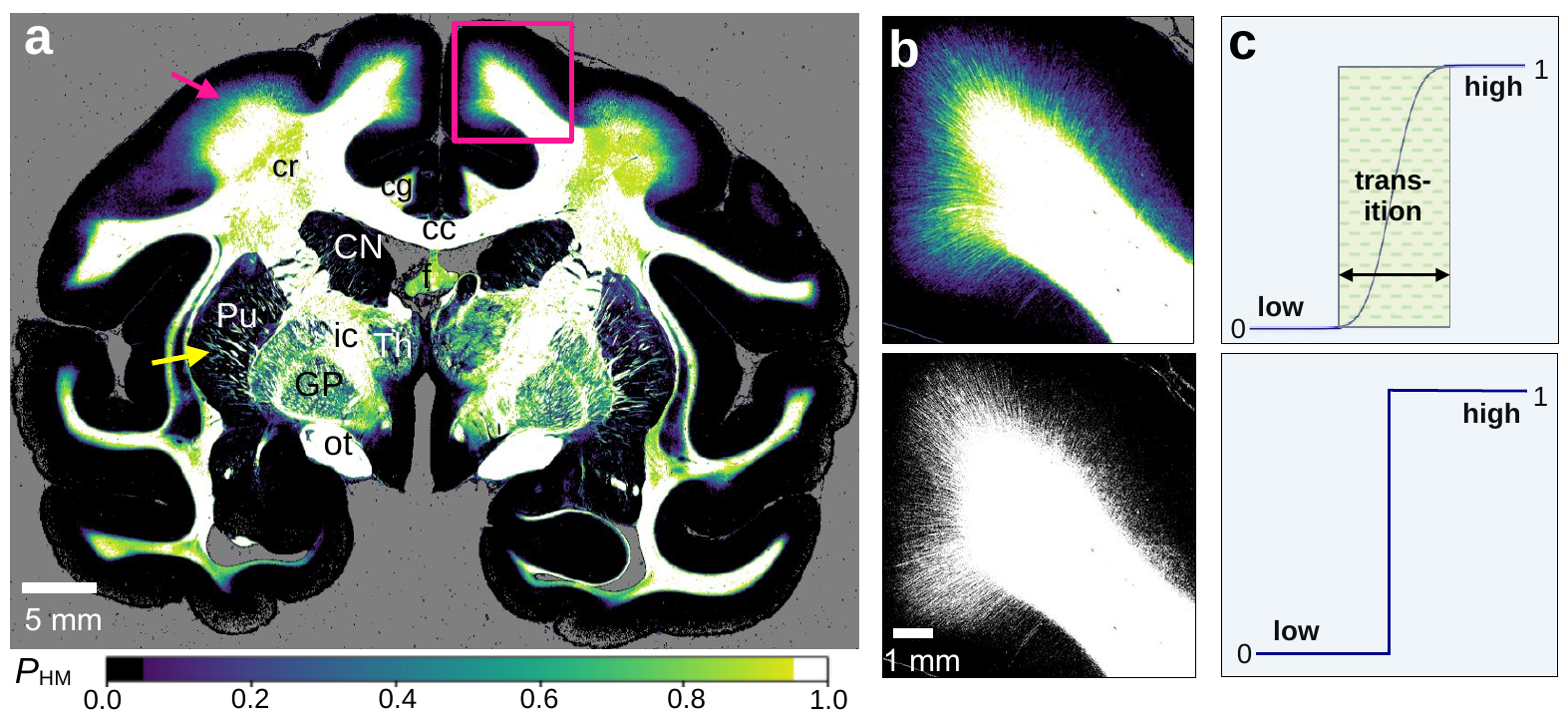}
	\caption{Myelination maps of the coronal vervet brain section. (\textbf{a}) HM-probability map of the whole section (computed with 25\,\% sample size and 200 iterations). Regions with $\Prob < 0.05$ are considered to have low myelination (LM-regions, in black), regions with $\Prob > 0.95$ are considered to have high myelination (HM-regions, in white). The transition zone is shown in different colours. The magenta and yellow arrow point to regions with a smooth and sharp transition, respectively. cc = corpus callosum, cg = cingulum, cr = corona radiata, f = fornix, ic = internal capsule, ot = optic tract, CN = caudate nucleus, GP = globus pallidus, Pu = putamen, Th = thalamus. (\textbf{b}) Enlarged view of the region marked by the magenta rectangle in (a). The top image shows the HM-probability map, the bottom image shows a sharp separation between HM/LM-regions without any transition zone. (\textbf{c}) Sketch of the functions used to describe the transition zone between HM/LM-regions. In the HM-probability map, a \textcolor{black}{sigmoid} function is used to describe the transition zone (top), the width depends on the computed mean of the threshold parameters, cf.\ \cref{fig:2}c. The bottom figure illustrates the binary classification. LM = 0, HM = 1.}
	\label{fig:3}
\end{figure}

\textcolor{black}{To realize a smooth transition between regions with low myelination ($\Prob = 0$) and high myelination ($\Prob = 1$),} the width of the transition zone \textcolor{black}{was} described by \textcolor{black}{a sigmoid} function (\textcolor{black}{cf.} top image in \cref{fig:3}c). 

The result is an \textit{HM-probability map} which indicates the probability $\Prob$ that a region is highly myelinated (HM-region)\textcolor{black}{. In this representation,} regions with $\Prob > 0.95$ were considered as HM-regions, regions with $\Prob < 0.05$ as LM-regions, and everything in between as transition zone. 
Depending on where the transmittance and retardation values of an image pixel lie within this transition zone, different values (between 0 and 1) were assigned.
The HM-probability map was computed using the following formula:
\begin{alignat}{2}
\Prob &= -0.5 \cdot \text{erf}
        \left(
            \cos\left(
                \frac{3}{4}\pi - \arctan2\left(
                    \frac{
                        \Delta \IT
                    }{
                        \Delta \textcolor{black}{R}
                    }
                \right)
        \right) \cdot \sqrt{
            \left(
                \Delta \IT
            \right)^{2} + 
            \left(
                \Delta \textcolor{black}{R}
            \right)^{2}
        }
        \right),
\end{alignat}
where \textcolor{black}{erf is the error function and}:
\begin{alignat}{2}
\Delta \textcolor{black}{R} &\equiv 
\begin{cases}
    \left|\frac{\textcolor{black}{R} - \rthres}{\textcolor{black}{R_{\text{thres}+}}}\right|\,, \quad \textcolor{black}{R} - \rthres &> 0, \\[5pt]
    \left|\frac{\textcolor{black}{R} - \rthres}{\textcolor{black}{R_{\text{thres}-}}}\right|\,, \quad \textcolor{black}{R} - \rthres &\leq 0, \\
\end{cases} \\
\Delta \IT &\equiv 
\begin{cases}
    \left|\frac{\IT - \Ithres}{\textcolor{black}{T_{\text{thres}+}}}\right|\,, \quad  \IT - \Ithres &> 0, \\[5pt]
    \left|\frac{\IT - \Ithres}{\textcolor{black}{T_{\text{thres}-}}}\right|\,, \quad  \IT - \Ithres &\leq 0. \\
\end{cases}
\end{alignat}

\Cref{fig:3}a shows the HM-probability map of the coronal vervet monkey brain section (HM-regions in white, LM-regions in black, transition zone in different colours). \Cref{fig:3}b shows a zoom-in of a transition region (top) in direct comparison to the corresponding binary classification of HM/LM-regions (bottom).
While there is a sharp separation between fibre bundles and surrounding tissue in the putamen (yellow arrow), there is a smooth transition for fibres spreading from the white matter into the cortex (magenta arrow).
The slightly lower $\Prob$-values in the corona radiata (cr), which is a region rich of white matter tracts, are due to the steep and crossing nerve fibres which impair the classification of regions (see Discussion).


\subsection*{\textcolor{black}{Automated computation of fibre inclinations}}

\textcolor{black}{Before \cref{eq:final} can be used to compute the nerve fibre inclinations with the $\Prob$ values obtained from above, the other parameters in the formula need to be determined: $\{\rmax, \IM, \ic\} \in [0,1]$. Also, the transmittance values $T$ need to be adjusted before computing the fibre inclinations.}


\paragraph{Computation of model parameters.}

\begin{figure}[H]
	\centering
	\includegraphics[width=0.7\textwidth]{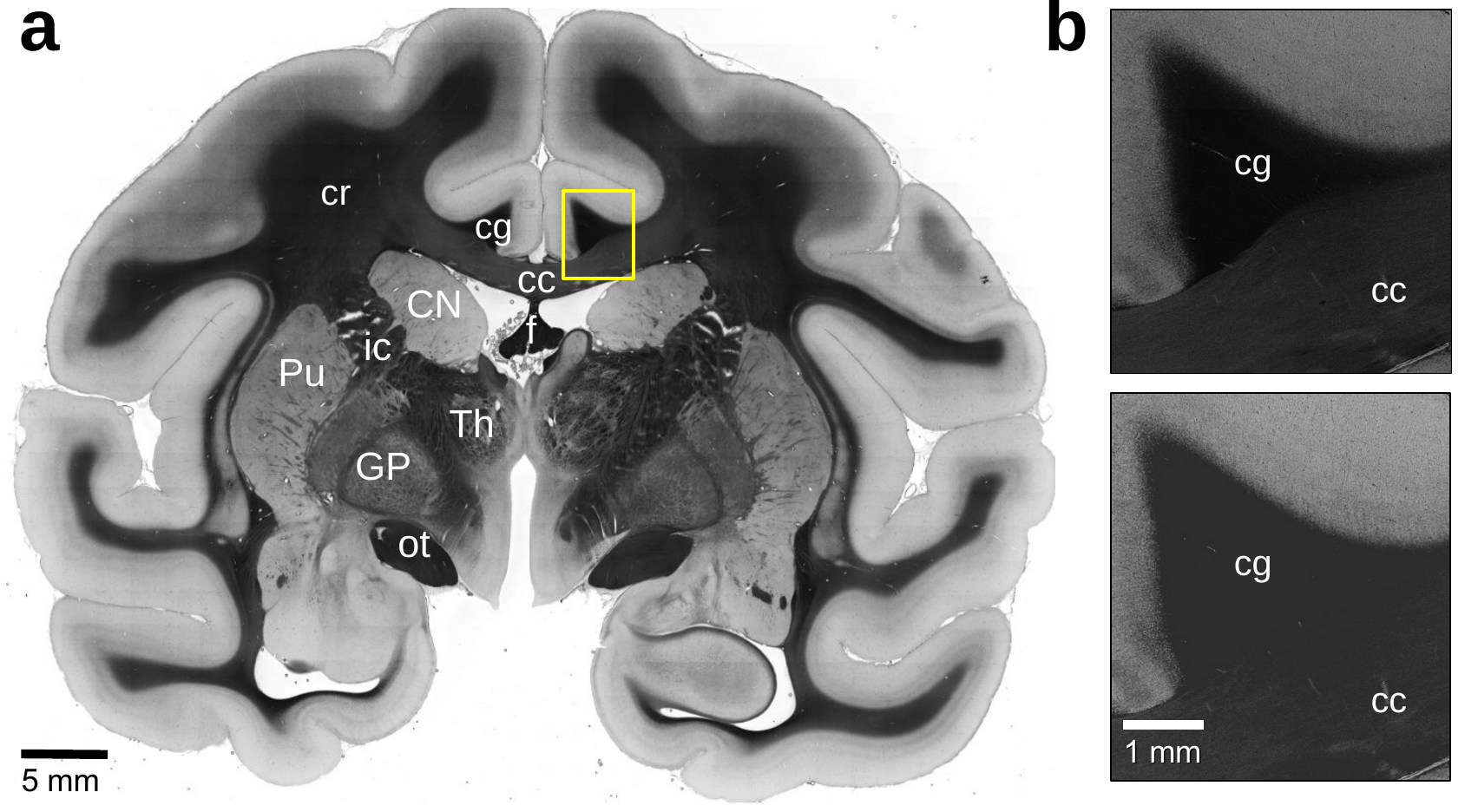}
	\caption{Correction of transmittance values. (\textbf{a}) Median-filtered transmittance image of the coronal vervet brain section\textcolor{black}{.} Anatomical regions are labelled for better reference (refer to \cref{fig:3}).
	(\textbf{b}) Comparison of transmittance images \textcolor{black}{(}yellow rectangle) before (top) and after (bottom) shifting the transmittance values of out-of-plane nerve fibres to $\Irmax$\textcolor{black}{.}}
	\label{fig:4}
\end{figure}

As the transmittance \textcolor{black}{values are} used to correct the fibre inclinations in the HM-regions (cf.\ \cref{eq:alpha2}), cellular structures in the transmittance image might lead to artifacts in the resulting inclination image. Therefore, \textcolor{black}{a circular median filter with five pixel radius was applied to the transmittance image beforehand.} To retain sharp borders between HM-and LM-regions, the median filter was applied to HM-and LM-regions \textcolor{black}{individually (ignoring background pixels)}. \Cref{fig:4}a shows the median-filtered transmittance image of the coronal vervet brain section\textcolor{black}{.}

To avoid the saturation of image pixels ($\alpha=0^{\circ}$ or $90^{\circ}$), the \textcolor{black}{highest retardation values were} computed for HM-and LM-regions separately, yielding $\rmaxHM$ and $\rmaxLM$. 
The highest retardation value in LM-regions is closely related to $\rthres$, which \textcolor{black}{was used to separate} HM-and LM-regions in the retardation histogram. Hence, $\rmaxLM$ was computed similar to $\rthres$, as the point of maximum curvature in the retardation histogram, considering only image pixels in the LM-regions ($\Prob < 0.05$).
The parameter $\rmaxHM$ was calculated
\textcolor{black}{by considering the retardation values in the HM-regions ($\Prob \geq 0.95$): From the image pixels with the highest retardation values (connected pixels, accounting for 0.009--0.011\,\% of the HM-regions), the} average of the highest 10\,\% was used as $\rmaxHM$.
The average transmittance of all selected pixels was used as $\IM$. The mode of the transmittance values in the LM-regions ($\Prob < 0.05$) was used as $\ic$.

As out-of-plane nerve fibres have lower transmittance values than in-plane nerve fibres due to an increased amount of scattering (see Menzel \ea\cite{menzel2020}), all (median-filtered) transmittance values smaller than $\IM$ (transmittance of the region with \textcolor{black}{the highest retardation values, expected to contain densely packed nerve fibres}) were \textcolor{black}{clipped} to this value, so that $\IT \geq \IM$.

\Cref{fig:4}b shows the median-filtered transmittance image (enlarged view of the region marked by the yellow rectangle) without this correction (top) and with correction (bottom). Without correction, the transmittance values of the cingulum (cg), which contains mostly very steep out-of-plane nerve fibres, are notably darker than those of the corpus callosum (cc), which contains mostly in-plane nerve fibres. The same applies to the fornix (f) and the optic tract (ot).


\paragraph{Software.}
The HM-probability maps and the nerve fibre inclinations were computed with the specifically developed open-source software \textit{PLImig}. \textcolor{black}{The algorithm is parallelised and uses GPU acceleration, allowing for a fast and reproducible computation of nerve fibre inclinations in large brain sections.} For more information, the reader is referred to our GitHub page (\url{https://github.com/3d-pli/PLImig}). 
Depending on the available graphics memory, the images were processed in a different number of chunks to avoid overflow. 
The coronal and sagittal rat brain sections (417 MB and 843 MB, using 32-bit float values) \textcolor{black}{were computed using an NVIDIA RTX 3070 and 8 GB of RAM. The coronal and sagittal vervet brain sections (6 GB and 6.5 GB) were processed on the supercomputer JURECA\cite{jureca}.}
Using 512 GB RAM, one GPU (NVIDIA A100 40GB) and 128 CPU cores on one node, the computing time for the coronal vervet brain section (34\,669 $\times$ 46\,341 pixels) was 15 min for the HM-probability map and 5 min for the rest.


\section*{Results}


\subsection*{\textcolor{black}{Comparison of different models for computing the fibre inclination angle in 3D-PLI}}

\begin{figure}[ht]
	\centering
	\includegraphics[width=\textwidth]{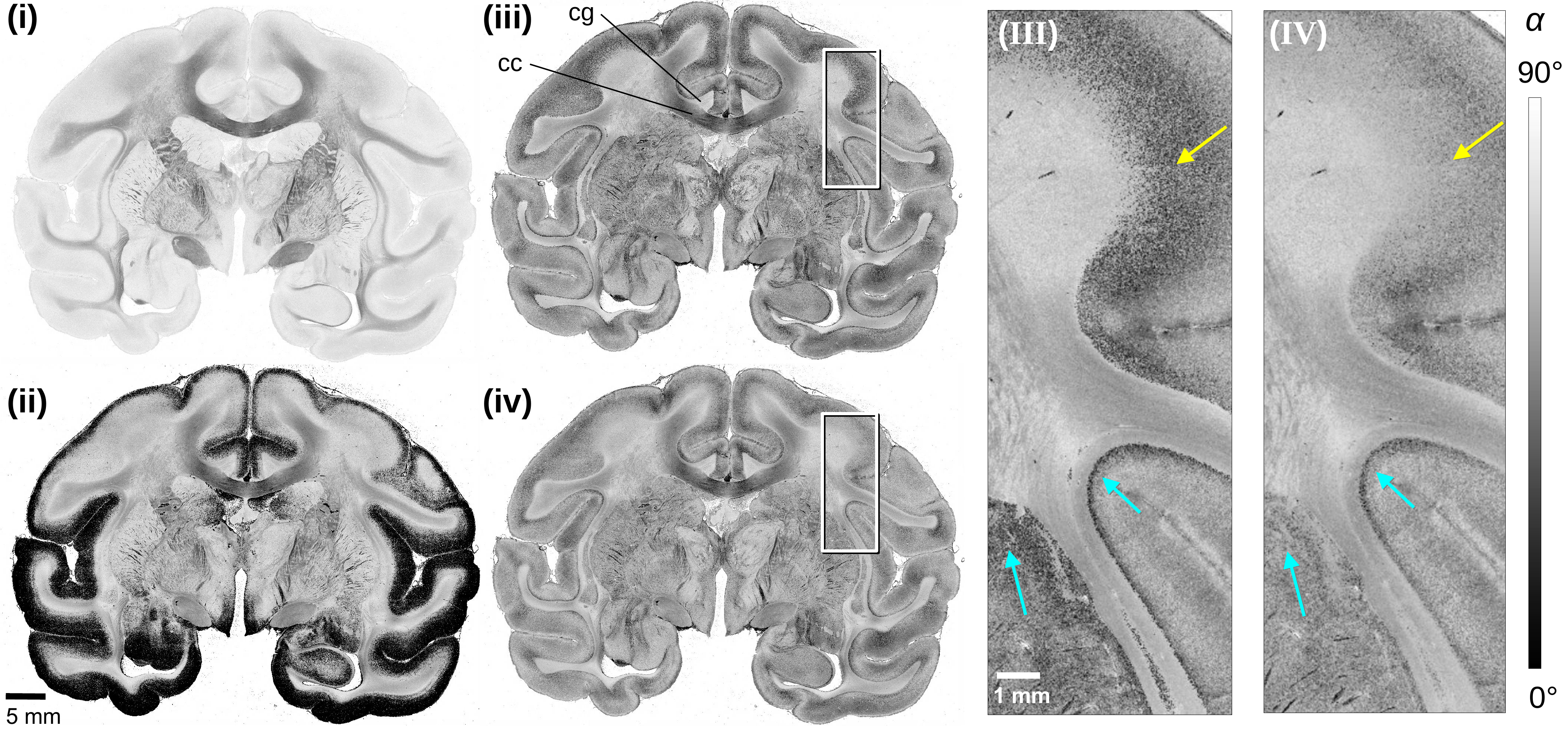}
	\caption{Inclination maps of the coronal vervet brain section computed with different models (\cref{eq:alpha,eq:alpha2,eq:final}):
	(i) unweighted model ($\alpha_{\text{LM}}$) applied to whole section, (ii) transmittance-weighted model ($\alpha_{\text{HM}}$) applied to whole section, (iii) both models applied separately to LM- and HM-regions, (iv) linear transition between both models depending on the computed HM-probability value. The images on the right ((III) and (IV)) show enlarged views of the regions marked by the white rectangles in (iii) and (iv), respectively. cc = corpus callosum, cg = cingulum. The yellow and cyan arrowheads point to two different types of transition zones located between LM- and HM-regions.}
	\label{fig:5}
\end{figure}

To illustrate the difference between the unweighted model and the transmittance-weighted model for computing the fibre inclination (\cref{eq:alpha,eq:alpha2}) as well as the effect of the transition zone between HM-and LM-regions (\cref{eq:final}), the fibre inclinations were computed for \textcolor{black}{four different scenarios} (see \cref{fig:5}): (i) unweighted model ($\alpha_{\text{LM}}$) applied to whole brain section, (ii) transmittance-weighted model ($\alpha_{\text{HM}}$) applied to whole brain section, (iii) unweighted and transmittance-weighted models applied to LM- and HM-regions \textcolor{black}{individually}, and (iv) unweighted and transmittance-weighted models applied to LM- and HM-regions using the HM-probability map combining both formulas in the transition zone.

When applying the unweighted model to the whole image, using the \textcolor{black}{highest} retardation of the brain section as reference ($\rmaxLM=\rmax$), the fibre inclinations in \textcolor{black}{low myelinated regions (\eg cortex)} are highly over-estimated (\cref{fig:5}(i)). On the other hand, when applying the transmittance-weighted model to the whole image ($\rmaxHM=\rmax$), the fibre inclinations in \textcolor{black}{these regions} are under-estimated, generating a lot of saturated values at $\alpha=0^{\circ}$ in black (\cref{fig:5}(ii)).
When applying the two different models to HM-and LM-regions \textcolor{black}{individually}, using their \textcolor{black}{highest retardation values} as reference ($\rmaxHM$, $\rmaxLM$), the estimated fibre inclinations become more realistic throughout the entire brain section. The contrast between out-of-plane fibres (cingulum = cg) and in-plane fibres, both in regions with densely packed nerve fibres (corpus callosum = cc) and in regions with bulked fibre architecture (cortex), becomes clearly visible (\cref{fig:5}(iii)). However, the sharp separation of HM- and LM-regions leads to an underestimation of the inclination (dark pixels) in regions where fibres extend to cortical areas (see yellow arrow in (III)), even worsening where sharp bending is involved (cyan arrows in (III)). The fourth model using the HM-probability map to combine the formulas for HM- and LM-regions in the transition zone reduces the underestimation considerably, leading to a more continuous course of fibres from deep white matter regions into the cortex (see yellow and cyan arrows in (IV))\textcolor{black}{, corresponding to the anatomical expectation}. 

\subsection*{\textcolor{black}{Comparison of fibre inclinations obtained from 3D-PLI and TPFM}}

To quantitatively evaluate the automatically computed nerve fibre inclinations, the caudate putamen of a coronal rat brain section was measured both with 3D-PLI and with two-photon fluorescence microscopy (TPFM) \textcolor{black}{as described in the Methods}.

\begin{figure}[ht]
	\centering
	\includegraphics[width=0.7\textwidth]{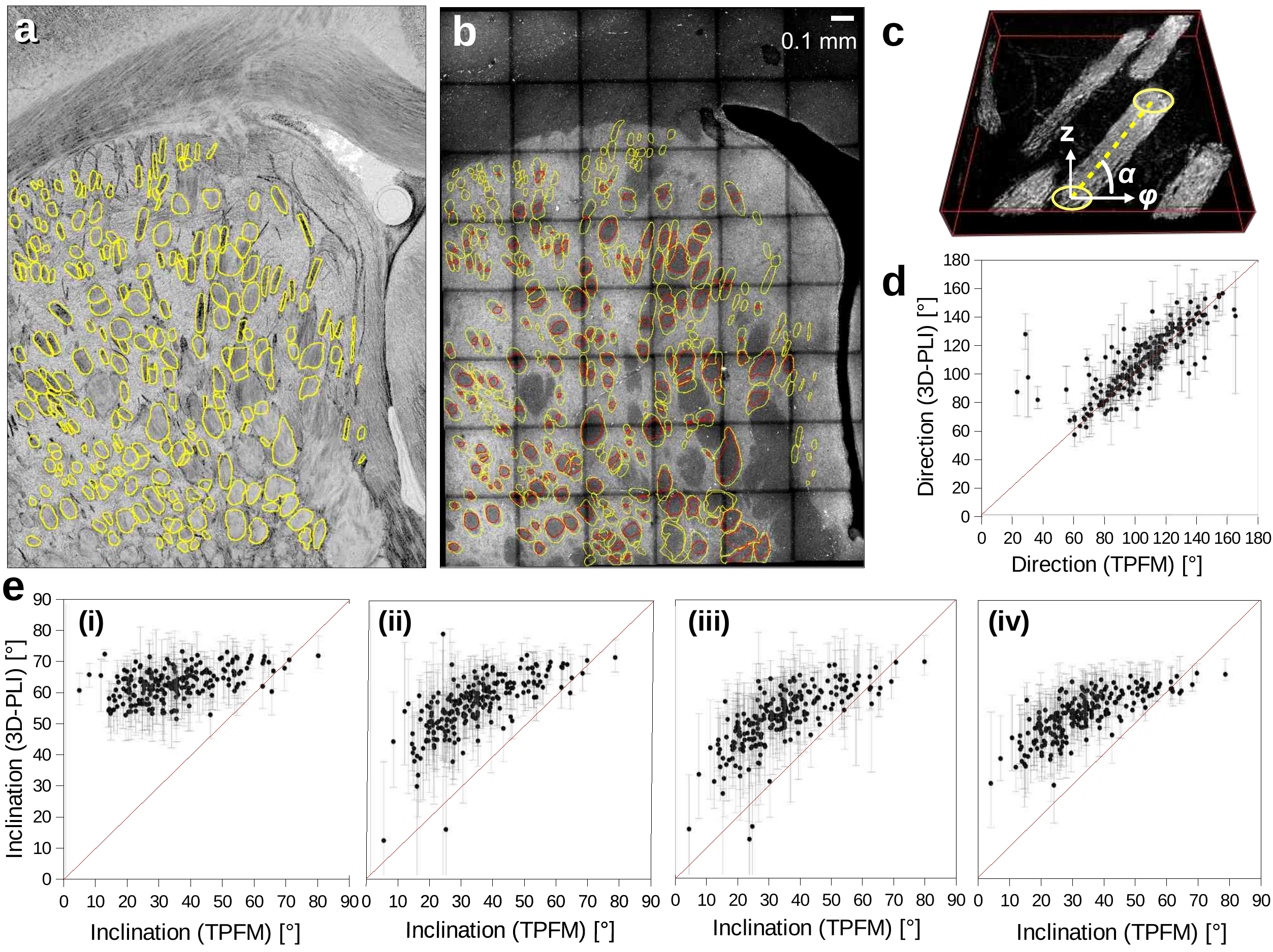}
	\caption{Comparison of 3D-PLI and TPFM nerve fibre orientations for a coronal rat brain section (230 selected fibre bundles in the caudate putamen).
(\textbf{a}) Inclination map obtained from the 3D-PLI measurement (using \cref{eq:final}). The fibre bundles selected for evaluation are labelled by yellow lines. (\textbf{b}) Maximum intensity projection of the TPFM image stack (containing $6 \times 8$ tiles). The yellow contours show the first and the last fully visible cross-sections of the fibre bundles in the TPFM image stack which were used to geometrically reconstruct the fibre inclinations. The red shapes denote the overlap of the cross-sections. \textcolor{black}{Only fiber bundles with defined orientation (not crossing or bending fibers) were selected for evaluation.} (\textbf{c}) 3D-model of the TPFM image stack \textcolor{black}{showing only fibrous tissue} (1 tile, adapted from Menzel \ea\cite{menzel2020}, Fig.\ S2c). The three-dimensional fibre orientation (dashed yellow line, defined by the in-plane direction angle $\varphi$ and the out-of-plane inclination angle $\alpha$) was geometrically computed from the centres of gravity of the bundle cross-sections mentioned above (yellow shapes). (\textbf{d}) Direction angles of the fibre bundles derived from the aligned 3D-PLI and TPFM measurements. The grey bars denote the positive and negative standard deviation when averaging the pixel values over the selected regions in the 3D-PLI direction image. (\textbf{e}) Inclination angles of the fibre bundles derived from the aligned 3D-PLI and TPFM measurements. The 3D-PLI inclinations were computed for four different scenarios (cf.\ \cref{fig:5}): (i) unweighted model ($\alpha_{\text{LM}}$) applied to whole brain section, (ii) transmittance-weighted model ($\alpha_{\text{HM}}$) applied to whole brain section, (iii) unweighted and transmittance-weighted models applied to LM- and HM-regions \textcolor{black}{individually}, (iv) unweighted and transmittance-weighted models applied to LM- and HM-regions using a linear combination of both formulas in the transition zone.}
	\label{fig:6}
\end{figure}

\textcolor{black}{As} TPFM allows an in-depth scan of the sample\textcolor{black}{, it allows to} directly assess the three-dimensional course of nerve fibre bundles in the caudate putamen (cf.\ \cref{fig:6}c), so that it can serve as a comparison for the nerve fibre orientations determined by 3D-PLI. 230 fibre bundles with well-defined shape and orientation were selected for evaluation (\cref{fig:6}a,b). The average 3D-PLI inclination of each bundle was plotted against the TPFM inclination, which was geometrically reconstructed from the corresponding bundle \textcolor{black}{(see Methods)}. \Cref{fig:6}d,e show the results for the in-plane fibre direction angle $\varphi$ and the out-of-plane fibre inclination angle $\alpha$ for the four different scenarios displayed in \cref{fig:5}.

While the direction angles computed from 3D-PLI correspond mostly to those obtained from TPFM (see \cref{fig:6}d), the computed 3D-PLI inclination angles are generally larger than those obtained from TPFM (see \cref{fig:6}e). When applying the unweighted model to the whole brain section (i), the computed 3D-PLI inclinations are highly over-estimated, especially for lower fibre inclinations ($< 50^{\circ}$). The computed 3D-PLI inclinations range between $50^{\circ}$ and $70^{\circ}$, mostly independent of the actual inclination of the nerve fibres. 
When applying the transmittance-weighted model to the whole brain section (ii), 
the computed 3D-PLI inclinations become much more similar to those obtained from TPFM. However, even when taking their standard deviations ($\Delta\alpha \approx 30^{\circ}$) into account, most values do not reach the TPFM inclinations.
When applying the unweighted and transmittance-weighted models \textcolor{black}{individually} to LM- and HM-regions regions (iii), a global shift of the 3D-PLI inclinations of about $-5^{\circ}$ can be observed.
When using a linear transition between the models (iv), the computed 3D-PLI inclinations are still over-estimated, but the difference to the fibre inclinations derived from the TPFM measurements becomes slightly smaller at higher inclinations.

\subsection*{\textcolor{black}{Fibre inclinations} in coronal and sagittal brain sections}

\Cref{fig:7} shows the HM-probability maps and the corresponding inclination maps (computed from \cref{eq:final}) for coronal and sagittal sections of vervet monkey and rat brains.

The HM-probability maps reliably separate regions with low myelination (black) from the rest. 
Individual nerve fibres are visible in the putamen (Pu) of the coronal vervet brain section, and in the caudate putamen (CPu) of the coronal and sagittal rat brain sections.
The HM-probability values in the white matter are slightly reduced (light green) for in-plane crossing fibres (cr) and steep out-of-plane fibres (coronal: cg, f; sagittal: cc), yielding slightly lower estimations of the inclination angles (cf.\ \cref{fig:6}e). 
The inclination maps show a good distinction of regions with in-plane and out-of-plane nerve fibres.
Although the coronal and sagittal sections belong to different brains, the different section planes (coronal/sagittal) clearly show an inverse pattern: In the coronal section plane, the computed inclination angles of the corpus callosum (cc) are much lower than those in the cingulum (cg) or the fornix (f). In the sagittal section plane, it is exactly the other way around.


\section*{Discussion}

When analysing measurement data from 3D-polarised light imaging (3D-PLI) of histological brain sections,
the determination of the out-of-plane nerve fibre inclinations has been a major challenge, requiring anatomical knowledge and effortful manual adjustment of parameters.
Here, we introduced an automatic computation of the fibre inclinations, by evaluating the retardation and transmittance images (normalised amplitude and average of the measured 3D-PLI signal). 
As these images are generated during standard 3D-PLI measurements, past measurements can be easily evaluated to improve the interpretation of fibre orientations in 3D-PLI data. In particular, no special equipment like a tilting stage is required to compute the nerve fibre inclinations. In contrast to the manual adjustment, no expert knowledge is required, the results for each measured brain section are reproducible, and large data sets can be analysed. In this way, we were able to provide a reliable estimate of the nerve fibre inclinations, which can still be adjusted by an expert in a second step.

In addition to the fibre inclinations, we provided an HM-probability map, indicating the probability that a region is highly myelinated. 
Benning \ea\cite{benning2021} proposed a binary separation of regions with low and high myelination. We improved the existing algorithm (taking different features/plateaus of the histograms into account and excluding background tissue when determining the threshold parameters). Furthermore, we introduced the important concept of transition zones.
The classification of brain regions into regions with high myelination (HM-regions), low myelination (LM-regions), and transition zones is not only relevant when applying different inclination formulas. 
Another possible application might be Independent Component Analysis (ICA), which has been used for noise and artefact removal in 3D-PLI images\cite{dammers2010}, especially in regions with low myelination (low birefringence signals)\cite{benning2021}. The HM-probability map allows to determine different degrees of myelination, helping to improve the ICA in those regions and enabling enhanced tissue segmentation.

The HM-probability maps and fibre inclinations were computed with a specifically developed open-source software, which is parallelised and uses GPU acceleration to efficiently process large images. While the computation of the fibre inclinations takes less than 5 min for the coronal vervet brain section (34\,669 $\times$ 46\,341 px), the generation of the HM-probability map takes about 15\,min. \textcolor{black}{The sampling consumes a major part of the computing time, but it is much more efficient than alternative approaches, e.g. defining transition zones by the local variance of pixel values (requiring to study many surrounding pixels).} 

\begin{figure}[H]
	\centering
	\includegraphics[width=0.8\textwidth]{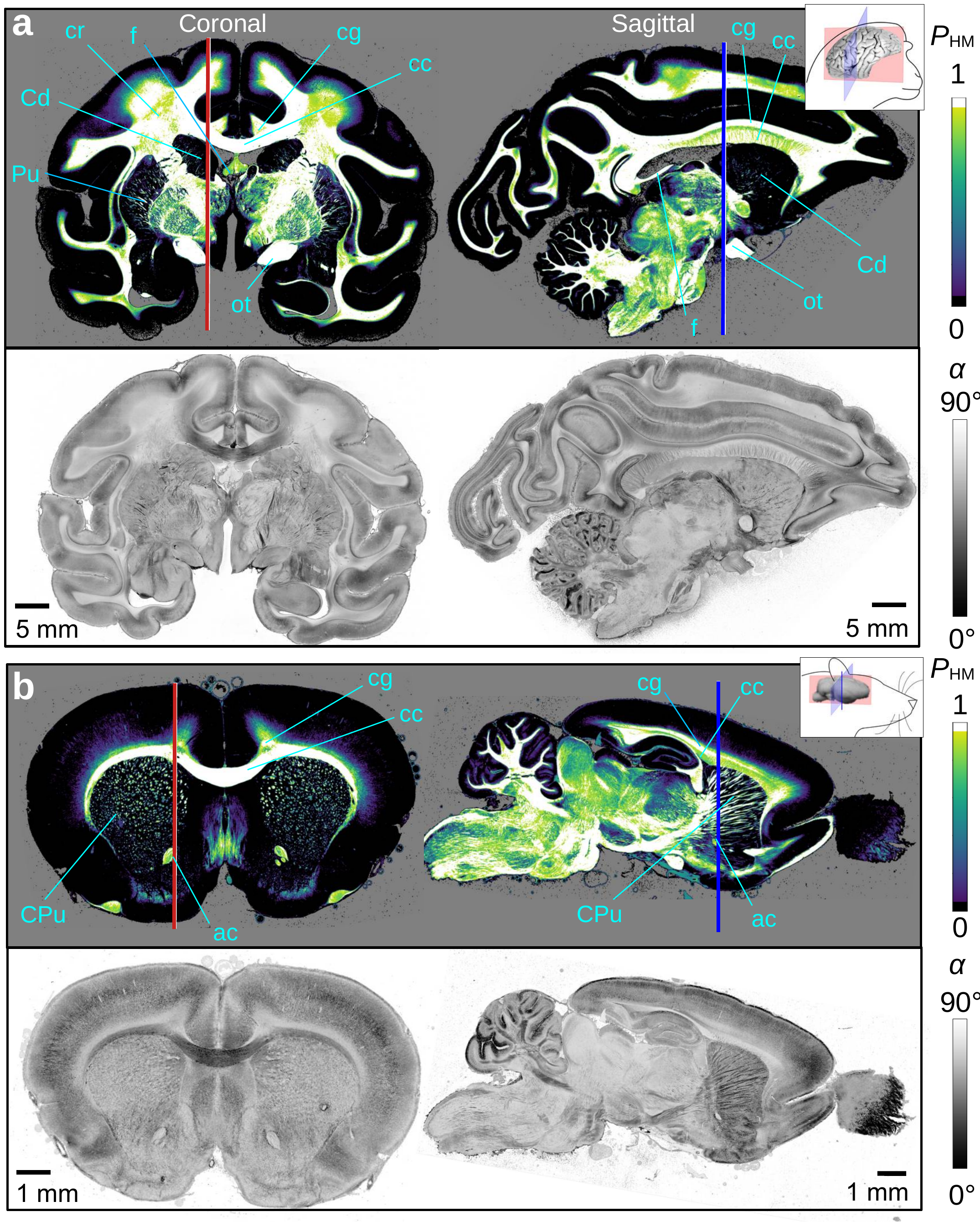}
	\caption{HM-probability and inclination maps of coronal and sagittal sections of vervet monkey brains (a) and rat brains (b).
	The upper images show the HM-probability maps $\Prob$ computed from 200 iterations and 25\,\% sample size. The lower images show the inclination maps $\alpha$ computed from the linearly combined formulas of LM- and HM-regions (\cref{eq:final}), using the HM-probability value. The red and blue vertical lines indicate the approximate position of the sagittal and coronal section plane in the complementary brain section. Selected anatomical regions are labelled for reference (ac = anterior commissure, cc = corpus callosum, cg = cingulum, cr = corona radiata, f = fornix, ot = optic tract, Cd = caudate nucleus, CPu = caudate putamen, Pu = putamen).}
	\label{fig:7}
\end{figure}

The choice of the reduced sample size and number of iterations (cf.\ \cref{fig:2}c) has a large impact on the computing time. When using 100\,\% instead of 25\,\% sample size, four times more computing time (\ie more than one hour) is required. To further reduce computing time, the number of iterations can be reduced to 100 without noticeable deviation of the threshold parameters from their asymptotic optimum (cf.\ \cref{fig:2}c).
Also, a more advanced bootstrapping approach (requiring smaller sample sizes) could be employed. 
The program could be extended to run on multiple GPUs, using the chunking algorithm that is already present. This would allow to process even larger brain sections and ideally achieve a linear speed-up.

While the software was shown to yield reliable results and correctly reproduce the overall inclination of major nerve fibre structures, several aspects should to be taken into account. 
First of all, the transmittance-weighted model used to compute the fibre inclinations in the HM-regions depends on the transmittance values. When the transmittance image has not enough contrast (\eg in brain sections measured several days/weeks after tissue embedding) or when the transmittance image shows irregularities (caused \eg by an inhomogeneous distribution of the embedding glycerine solution or tile/stitching artefacts), this leads to artefacts in the resulting inclination map, which would not be visible when solely using the birefringence signal (retardation) for computing the inclination, as in the LM-regions. Hence, the software should only be used when the transmittance image contrast is free of artefacts.
Another point to take into account is that the algorithm uses the transmittance and retardation histograms to compute \textcolor{black}{the parameters} in the inclination formula. A reliable computation of parameters only works when the histograms include a broad variety of tissue components
\textcolor{black}{as} well as various nerve fibre structures (in-plane and out-of-plane fibres).

While the computed inclinations can serve as an estimate of the real out-of-plane angle in homogeneous nerve fibre tissue, they are highly over-estimated in regions with in-plane crossing nerve fibres due to the reduced birefringence signal\cite{dohmen2015}. As shown in Menzel \ea\cite{menzel2020}, the transmittance image can serve as an indicator whether large inclination angles belong to out-of-plane nerve fibres or are caused by in-plane crossing fibres or a low fibre density. Regions with high inclination values can be considered as reliable if the transmittance in these regions is notably lower than in regions with in-plane nerve fibres. When comparing the transmittance image (\cref{fig:4}a) with the inclination image (\cref{fig:7}a) for the coronal vervet brain section, it becomes apparent that cingulum and fornix fulfill this criterion (\ie their transmittance is notably lower than the transmittance of the corpus callosum which contains mostly in-plane nerve fibres).
The corona radiata is dotted with regions that show substantially darker transmittance values than the corpus callosum (caused by the superior longitudinal fascicle -- a steep fibre tract crossing this region), but it also contains regions with similar transmittance values, indicating that the high inclination values in these areas cannot be considered as reliable and are most probably caused by in-plane crossing nerve fibres.
As the threshold parameters in the transmittance histogram are very susceptible to small changes, we decided to keep an automated detection of crossing nerve fibres for future algorithms.

So far, the nerve fibre inclinations were computed by applying a transmittance-weighted model to the whole brain section and manually adjusting the parameters.
The models presented here (with and without transmittance weighting in HM- and LM-regions, \cref{eq:alpha,eq:alpha2}) lead to a major enhancement of the computed fibre inclinations in \textcolor{black}{regions with low myelination} (cf.\ \cref{fig:5}(ii) and (iii)). The computation of an HM-probability map allowed for the first time to consider transition zones, reducing step artefacts at the boundaries between HM- and LM-regions (cf.\ \cref{fig:5}(III) and (IV)): 
The course of nerve fibres from 
\textcolor{black}{highly myelinated regions}
into the cortex becomes much more continuous (yellow arrows), even border regions with sharp bending that were clearly visible as dark stripes become more softened (cyan arrows). However, the dark fissures of the inclination at the borders are still present in some regions and still need to be improved by another tuning of \textcolor{black}{$R_{\text{thres+/-}}$ and $T_{\text{thres+/-}}$}.
As shown in \cref{fig:7}, the consistency of nerve fibre inclinations in whole coronal and sagittal brain sections of different species (monkeys and rodents) validates the reliability of the algorithms applied.

The findings presented in \cref{fig:6} show that the different models have different impacts on the computed inclinations of nerve fibre bundles (in the caudate putamen of a rat brain). 
When using the unweighted model (i), the computed inclinations are mostly independent of the real underlying fibre inclination. The transmittance-weighted model (ii) introduces the proportionality of the computed inclinations (they increase with increasing fibre inclination) in particular for values below $50^{\circ}$. The introduction of the HM-probability map (iv), necessary to generate valid fibre inclinations in \textcolor{black}{regions with low myelination} and to obtain a smooth transition at the boundaries, has only a small impact on the computed fibre inclinations of the nerve fibre bundles: (iv) almost shows a similar behaviour as (ii). Hence, the linear combination of both models (\cref{eq:final}) to compute the nerve fibre inclinations in regions with different degrees of myelination can be used without changing the inclination values in
\textcolor{black}{highly myelinated regions}.
Although the transmittance-weighted inclinations (ii-iv) are much more realistic than those computed with the unweighted model (i), the comparison of 3D-PLI and two-photon fluorescence microscopy (TPFM) shows a strong over-estimation of the computed fibre inclinations ($\Delta\alpha \approx 30^{\circ}$). However, the inclinations obtained from 3D-PLI might also be larger than those obtained from TPFM because the TPFM measurements have been performed several weeks afterwards so that the thickness of the brain section might have decreased due to dehydration. For this reason, a direct comparison of 3D-PLI and TPFM values is critical. However, as the affine image registration of 3D-PLI to TPFM measurements did not reveal large tissue deformations, maintaining also a high affinity of the corresponding direction angles, major deviations of the fibre orientations are unlikely.

In conclusion, we have introduced a versatile software that allows for the first time to automatically compute the out-of-plane angles of nerve fibres obtained from standard 3D-PLI measurements of brain tissue (without tilting). So far, these nerve fibre inclinations were computed by applying a transmittance-weighted model to the whole brain section, requiring anatomical knowledge and effortful manual adjustment of parameters. 
By distinguishing areas with different degrees of myelination and defining transition zones, we were able to apply a regionally specific computation of fibre inclinations \textcolor{black}{that} accounts for subtle changes in the tissue composition.
The developed algorithm is parallelised and uses GPU acceleration, allowing for a fast and reproducible computation of nerve fibre inclinations in large brain sections, thus greatly improving the analysis and interpretation of 3D-PLI data also for past measurements.


\paragraph{Code availability.}
The software \textit{PLImig} that was used to compute the HM-probability maps and the nerve fibre inclinations in \cref{fig:3,fig:5,fig:7} is publicly available on GitHub (\url{https://github.com/3d-pli/PLImig}).
Data analysis was performed with the open-source software \textit{Fiji} (\url{https://fiji.sc/Fiji}).


\paragraph{Data availability.}
All data supporting the findings of this study are included in the provided figures.


\bibliography{BIBLIOGRAPHY}

\section*{Acknowledgements}
We thank the lab team of the INM-1 (Institute of Neuroscience and Medicine, Forschungszentrum Jülich GmbH, Germany) for preparing the brain sections, Philipp Schlömer for the generation of the transmittance and retardation images, the members of the Fibre Architecture group (INM-1) for assisting in the manual evaluation of the TPFM inclinations, and Karl Zilles and Roger Woods for collaboration in the vervet brain project (National Institutes of Health under Grant Agreements No.\ R01MH092311 and 5P40OD010965).
This work was funded by the European Union's Horizon 2020 Framework Programme for Research and Innovation under the Specific Grant Agreement No.\ 945539 (``Human Brain Project'' SGA3) and the Helmholtz Association portfolio theme \textit{`Supercomputing and Modeling for the Human Brain'}.
We gratefully acknowledge the computing time granted through JARA-HPC on the supercomputer \textit{JURECA} \cite{jureca} at Forschungszentrum Jülich (FZJ). 

\section*{Author contributions statement}
M.M.\ designed and supervised the research.
J.R.\ developed the software and analysis methods, and visualised the results.
M.M., J.R., D.G.\ and M.A.\ contributed to the analysis of the data.
M.M., J.R., D.G., M.A.\ and K.A.\ contributed to the interpretation of the data.
I.C.\ performed the TPFM measurements.
M.M.\ wrote the first version of the manuscript. D.G., J.R., M.A., and K.A.\ contributed to the manuscript.\\

\noindent\textbf{Competing interests:} The authors declare no competing interests.

\end{document}